\documentclass[preprintnumbers,amsmath,amssymb,prd,floatfix,12pt,superscriptaddress,nofootinbib]{revtex4}
\usepackage{graphicx}
\usepackage{epsfig}
\usepackage{bm}
\usepackage{amsfonts}

\begin{document}

\title{Interacting entropy-corrected new agegraphic K-essence, tachyon and
dilaton scalar field models in non-flat universe}
\author{M. Umar Farooq}
\email{mfarooq@camp.nust.edu.pk}
\affiliation{Center for Advanced Mathematics and Physics,\\
National University of Sciences and Technology, Rawalpindi, 46000, Pakistan}
\author{Muneer A. Rashid}
\affiliation{Center for Advanced Mathematics and Physics,\\
National University of Sciences and Technology, Rawalpindi, 46000, Pakistan}
\author{Mubasher Jamil}
\email{mjamil@camp.nust.edu.pk}
\affiliation{Center for Advanced Mathematics and Physics,\\
National University of Sciences and Technology, Rawalpindi, 46000, Pakistan}

\begin{abstract}\vspace*{1.5cm} \centerline{\bf Abstract} \vspace*{1cm}
We present the new agegraphic dark energy model with the help of the
quantum corrections to the entropy-area relation in the setup of
loop quantum gravity. Employing this new form of dark energy, we
investigate the model of interacting dark energy and derive its
equation of state (EoS). We study the correspondence between the
K-essence, tachyon and dilaton scalar fields with the interacting
entropy-corrected new agegraphic dark energy in the non-flat FRW
universe. Moreover, we reconstruct the corresponding scalar
potentials which describe the dynamics of the scalar field.
\end{abstract}

\maketitle
\newpage
\section{\textbf{Introduction}}

One of the most outstanding developments in the last decade is the discovery
that at the current era our universe is undergoing the phase of accelerating
expansion \cite{1}. \textquotedblleft Dark energy\textquotedblright ,\ an
unknown exotic vacuum energy responsible for propelling the universe, is one
of the deepest mysteries in astrophysics. The dark energy possesses negative
pressure $p<0$ and positive energy density $\rho >0$ which is related by the
equation of state $p=\omega \rho .$ Astrophysical data suggests that about
two-third of the critical energy is stored in the dark energy component
apart from dark matter which contains only one third of the critical energy
density. One possible source of this cosmic expansion can be explained by
the general theory of relativity with the cosmological constant $\Lambda .$
Although the cosmological constant is the most obvious choice, but it
suffers from coincidence problem and the fine-tuning problems \cite{2}. To
overcome these problems , several alternative models have been suggested;
among them are dynamical scalar field $\phi $ with suitably defined scalar
field potential $V(\phi )$ termed as quintessence \cite{3}, quintom \cite{4}%
, k-essence \cite{5}, tachyon \cite{6}, phantom \cite{7}, dilatonic ghost
condensate \cite{8} to name a few. In addition, there are other proposals on
dark energy such as interacting dark energy models \cite{9}, brane world
models \cite{10} and Chaplygin gas models \cite{11} etc.

In the last few years, the holographic dark energy (HDE) models
\cite{12} and agegraphic dark energy models \cite{13} have received
a considerable interest. Holographic principle is a speculative
conjecture about proposed quantum theories of gravities . According
to holographic principle, the information contained in a volume may
be described by a theory that lies on the boundary of that space
\cite{14}. In the study of thermodynamics of the black hole, there
is a maximum entropy in a box of size $L,$ known as the Bekenstein
Hawking entropy bound $S\backsim m_{p}^{2}L^{2}$ which scales as the
area of the box $A\backsim L^{2}.$ To avoid the breakdown of quantum
field theory in the frame work of quantum gravity, Cohen et al
\cite{15}
proposed that the entropy for an effective theory should satisfy $%
L^{3}\Lambda ^{3}\leq S^{3/4}\leq (m_{p}^{2}L)^{3/2}.$ Here $L$
associated with the size of a region which gives an infra-red
cut-off while $\Lambda $ corresponds to the ultra-violet cut-off.
The last expression can be transformed to $\rho _{\Lambda
}=3n^{2}m_{p}^{2}L^{2},$ where $3n^{2}$ is for convenience and
$m_{p}=1/\sqrt{8\pi G}$ is the Plank mass. In Einstein theory of
gravity, the definition of holographic dark energy requires the
entropy-area relationship $S\backsim A\backsim L^{2},$ where $A$ is
the area of the horizon. However in the context of loop quantum
gravity (LQG), this entropy-area relationship gets modified from the
inclusion of quantum effects. The quantum corrections provided to
the entropy-area relationship facilitates us to obtain correction in
the Hilbert action and vise versa
\cite{16}. The corrected entropy is%
\begin{equation}
S=\frac{A}{4G}+\tilde{\gamma}\ln (\frac{A}{4G})+\tilde{\beta},
\end{equation}%
where $\tilde{\gamma}$ and $\tilde{\beta}$ are constants of order unity. The
exact values of these parameters are unknown and still a matter of debate.
These corrections usually come into view in the black hole entropy in LQG
due to thermal and quantum fluctuations \cite{17}. The entropy-corrected
holographic dark energy (ECHDE) is given by \cite{18}%
\begin{equation}
\rho _{\Lambda }=3n^{2}m_{p}^{2}L^{-2}+\gamma L^{-4}\ln
(m_{p}^{2}L^{2})+\beta L^{-4},
\end{equation}%
where $\tilde{\gamma}$ and $\tilde{\beta}$ are dimensionless constants of
order unity. Clearly if we put $\tilde{\gamma}=\tilde{\beta}=0,$ we arrive
at the holographic dark energy model.

Though HDE is considered to be the most promising candidate for dark energy,
but there are some difficulties appear in HDE model. Choosing the future
event horizon of the universe as the length scale, the HDE model does not
contradict to the observed value of dark energy in the universe and can
propel the universe to an accelerated expansion phase. However, the current
properties of the dark energy determined by the future evolution of the
universe might violate causality. Moreover, it has been argued that this
model might be in contradiction to the age of the universe \cite{19}.

Recently, Cai \cite{13} has proposed a model, dubbed \textquotedblleft
agegraphic dark energy\textquotedblright\ (ADE)\ based on Karolyhazy
uncertainty relation $\delta t=\lambda t_{p}^{2/3}\tau ^{1/3}$\ \cite{20}, $%
\lambda $ is a numerical factor of order one and $t_{p}$ is the Plank time.
Following the Karolyhazy relation\ Maziashvili argued that the energy
density of \ metric fluctuation of Minkowski spacetime can be written as $%
\rho _{\Lambda }\backsim \frac{1}{t_{p}^{2}\tau ^{2}}\backsim \frac{m_{p}^{2}%
}{\tau ^{2}},$ where the time scale $\tau $ is chosen to be the age
of the universe $T=\int_{0}^{a}\frac{d\tilde{a}}{H\tilde{a}}$ and
the energy density of the agegraphic dark energy can be expressed as
$\rho _{\Lambda }=3n^{2}m_{p}^{2}T^{-2}$ \cite{13}. Since in the ADE
model the age of the universe is taken as the length measure instead
of the horizon distance, so the causality problem that appears in
the HDE model can be avoided. However, the ADE model might contain
an inconsistency \cite{21}. So to overcome this problem, soon after
the original ADE model, the authors \cite{22} proposed an
alternative model of dark energy so called the \textquotedblleft\
new agegraphic dark energy\textquotedblright\ (NADE) and its density
energy is
defined by%
\begin{equation}
\rho _{\Lambda }=3n^{2}m_{p}^{2}\eta ^{-2},
\end{equation}%
where $\eta $ is conformal time of the FRW-universe and is written as%
\begin{equation}
\eta =\int \frac{dt}{a}=\int_{0}^{a}\frac{da}{Ha^{2}}.
\end{equation}%
In this paper, we extend our study of NADE from the inclusion of
quantum corrections to the entropy-area relation. We investigate the
entropy-corrected version of the interacting NADE model so called
the entropy-corrected new agegraphic dark energy (ECNADE) and see
its effects in the non-flat universe. The plan of the work is as
follows. In Section 2, we study the ECNADE model in a non-flat
Friedmann-Robertson-Walker (FRW) universe and derive its equation of
state. In Section 3, we present a correspondence between the ECNADE
and the K-essence, tachyon and dilaton scalar fields. In each case,
we reconstruct the potential and dynamics for these scalar fields
which depict the accelerated expansion.

\section{\textbf{Interacting ECNADE model}}

We assume the background to be a spatially homogeneous and isotropic FRW
spacetime, given by%
\begin{equation}
ds^{2}=-dt^{2}+a^{2}(t)\left[ \frac{dr^{2}}{1-kr^{2}}+r^{2}(d\theta
^{2}+\sin ^{2}\theta d\varphi ^{2})\right] .
\end{equation}%
Here $a(t)$ is the dimensionless scale factor which is an arbitrary function
of time and $k$ is represents the curvature parameter which has dimensions
of length$^{-2}$. For the values $k=-1,0,1,$ the above metric represents the
spatially open, flat and closed FRW universe respectively. The first
Friedmann equation for the non-flat FRW-spacetime containing the dark energy
and dark matter is%
\begin{equation}
H^{2}+\frac{k}{a^{2}}=\frac{1}{3m_{p}^{2}}[\rho _{\Lambda }+\rho _{m}],
\end{equation}%
where $M_{p}^{2}=(8\pi G)^{-1}$ is modified Planck mass. Here
$H=\dot{a}/a$ is the Hubble constant while $\rho _{\Lambda }$ and
$\rho _{m}$ represent the energy densities of dark energy and matter
respectively. Let us define the
dimensionless energy density parameters as%
\begin{equation}
\Omega _{m}=\frac{\rho _{m}}{\rho _{cr}}=\frac{\rho _{m}}{3m_{p}^{2}H^{2}},%
\text{ }\Omega _{\Lambda }=\frac{\rho _{\Lambda }}{\rho
_{cr}}=\frac{\rho _{\Lambda }}{3m_{p}^{2}H^{2}},\text{ }\Omega
_{k}=\frac{k}{(aH)^{2}}.
\end{equation}%
With the help of these parameters, the Friedmann Eq. (6) takes the form%
\begin{equation}
1+\Omega _{k}=\Omega _{\Lambda }+\Omega _{m}.
\end{equation}%
In the setup of LQG, we would like to consider ECNADE whose length $L$ in
Eq. (2) is taken to be the conformal time $\eta $ given by (4). So the
density of the ECNADE can be written as%
\begin{equation}
\rho _{\Lambda }=3n^{2}m_{p}^{2}\eta ^{-2}+\gamma \eta ^{-4}\ln (m^{2}\eta
^{2})+\beta \eta ^{-4},
\end{equation}%
where $\gamma $ and $\beta $ are dimensionless constants of order one.
Employing the relationship $\Omega _{\Lambda }=\frac{\rho _{\Lambda }}{%
3H^{2}m_{p}^{2}},$ we get%
\begin{equation}
\Omega _{\Lambda } =\sqrt{\frac{3n^{2}m_{p}^{2}+\gamma \eta ^{-2}\ln
(m_{p}^{2}\eta ^{2})+\beta \eta ^{-2}}{3m_{p}^{2}H^{2}\eta^2}}.
\end{equation}%
Let us assume an interaction $Q=\Gamma \rho _{\Lambda }$ between ECNADE and
the cold dark matter (CDM) having $\omega _{m}=0$. The resulting energy
conservation equations for ECNADE and CDM are%
\begin{eqnarray}
\dot{\rho}_{\Lambda }+3H(\rho _{\Lambda }+p_{\Lambda }) &=&-Q, \\
\dot{\rho}_{m}+3H\rho _{m} &=&Q.
\end{eqnarray}%
Here overdot represents the differentiation with respect to cosmic time $t$.
These interacting models describe an energy flow between the dark energy and
dark matter so that no component shows energy conservation independently. If
two species are present in dominant form, they are definitely supposed to
interact. We choose $\Gamma =3b^{2}H(\frac{1+\Omega _{k}}{\Omega _{\Lambda }}%
)$ as the decay rate of the ECNADE component into CDM with a
coupling constant $b^{2}.$ The importance of interacting model also
emerges as it is good fit to the expansion history of the universe
as determined by the Supernovae and cosmic microwave background
\cite{1}.

Differentiate Eq. (9) with respect to time and using $\dot{\eta}=\frac{1}{a}$
and Eq. (10), we obtain%
\begin{equation}
\dot{\rho}_{\Lambda }=\frac{2\chi H}{a}\sqrt{\frac{3m_{p}^{2}\Omega _{\Lambda }}{%
3n^{2}m_{p}^{2}+\gamma \eta ^{-2}\ln (m_{p}^{2}\eta ^{2})+\beta \eta
^{-2}}},
\end{equation}%
where $\chi \equiv\gamma \eta ^{-4}-2\gamma \eta ^{-4}\ln
(m_{p}^{2}\eta ^{2})-3n^{2}m_{p}^{2}\eta ^{-2}-2\beta \eta ^{-4}$.
Making use of the above Eq. (13) in (11), we obtain the following
equation
of state parameter of the interacting ECNADE model%
\begin{equation}
\omega _{\Lambda
}=-1-\frac{2\chi}{3a}\Big(\frac{\sqrt{\frac{3m_{p}^{2}\Omega
_{\Lambda }}{3n^{2}m_{p}^{2}+\gamma \eta ^{-2}\ln (m_{p}^{2}\eta
^{2})+\beta \eta ^{-2}}}}{3n^{2}m_{p}^{2}\eta ^{-2}+\gamma \eta
^{-4}\ln (m_{p}^{2}\eta ^{2})+\beta \eta ^{-4}}\Big)
-b^{2}\Big(\frac{1+\Omega _{k}}{\Omega _{\Lambda }}\Big).
\end{equation}%

\section{\textbf{Correspondence between ECNADE and K-essence, tachyon and
dilaton scalar field}s}

The cosmological constant corresponds to a fluid with a constant
equation of state $\omega =-1$. Now, the observations which put
restriction on the value of $\omega $ to be close to that of
cosmological constant, explain bit less about time evolution of
$\omega $. So we need to consider a model in which the EoS of dark
energy evolves with time such as in inflationary cosmology. Scalar
field models arise in string theory and are studied as promising
candidates for dark energy. So far, ample literature dealing with
scalar field dark energy models is available. It includes
quintessence, K-essence, phantoms, tachyon, and dilaton among many.
In this section, we investigate the correspondence between the
interacting ECNADE model with the K-essence, tachyon and dilaton
scalar fields in the non-flat FRW universe. To illustrate this
correspondence, we first equate the interacting ECNADE density with
the corresponding scalar field density. Then we compare the equation
of state of the scalar field model with the EoS of the ECNADE.

\subsection{\textbf{Entropy-corrected new agegraphic K-essence model}}

The idea of the K-essence scalar field was motivated from the Born-Infeld
action of string theory \cite{23} and used as a source to explain the
mechanism for producing the late time acceleration of the universe. The
K-essence model is expressed by a general scalar field action which is
function of $\phi $ and $X$=$\dot{\phi}^{2}/2$ and is given by \cite{5}%
\begin{equation}
S=\int d^{4}x\sqrt{-g}\text{ }p(\phi ,X),
\end{equation}%
where the Lagrangian density $p(\phi ,X)$ corresponds to a pressure density
as%
\begin{equation}
p(\phi ,X)=f(\phi )(-X+X^{2}),
\end{equation}%
and the energy density of the field $\phi $ as%
\begin{equation}
\rho (\phi ,X)=f(\phi )(-X+3X^{2}).
\end{equation}%
The EoS parameter for the K-essence scalar field is obtained in the
following way%
\begin{equation}
\omega _{K}=\frac{p(\phi ,X)}{\rho (\phi ,X)}=\frac{X-1}{3X-1}.
\end{equation}%
After equating Eq. (18) with the ECNADE equation of state parameter (14), we
determine the expression for $X$ in the form%
\begin{equation}
X=\frac{2+\frac{2\chi}{3a}\frac{\sqrt{\frac{3m_{p}^{2}\Omega _{\Lambda }}{%
3n^{2}m_{p}^{2}+\gamma \eta ^{-2}\ln (m_{p}^{2}\eta ^{2})+\beta \eta ^{-2}}}%
}{3n^{2}m_{p}^{2}\eta ^{-2}+\gamma \eta ^{-4}\ln (m_{p}^{2}\eta ^{2})+\beta
\eta ^{-4}} +b^{2}(\frac{1+\Omega _{k}}{\Omega _{\Lambda }})}{4+%
\frac{2\chi}{a}\frac{\sqrt{\frac{3m_{p}^{2}\Omega _{\Lambda }}{%
3n^{2}m_{p}^{2}+\gamma \eta ^{-2}\ln (m_{p}^{2}\eta ^{2})+\beta \eta ^{-2}}}%
}{3n^{2}m_{p}^{2}\eta ^{-2}+\gamma \eta ^{-4}\ln (m_{p}^{2}\eta
^{2})+\beta \eta ^{-4}} +3b^{2}(\frac{1+\Omega _{k}}{\Omega
_{\Lambda }})}.
\end{equation}%
Using the above Eq. (19) and the relation $\dot{\phi}^{2}=2X,$ the
evolutionary form of the K-essence scalar field is determined to be%
\begin{equation}
\dot{\phi}=\left(
\frac{4+\frac{4\chi}{3a}\frac{\sqrt{\frac{3m_{p}^{2}\Omega _{\Lambda
}}{3n^{2}m_{p}^{2}+\gamma \eta ^{-2}\ln (m_{p}^{2}\eta ^{2})+\beta
\eta ^{-2}}}}{3n^{2}m_{p}^{2}\eta ^{-2}+\gamma \eta ^{-4}\ln
(m_{p}^{2}\eta ^{2})+\beta \eta ^{-4}}+2b^{2}(\frac{1+\Omega
_{k}}{\Omega
_{\Lambda }})}{4+\frac{2\chi}{a}\frac{\sqrt{\frac{3m_{p}^{2}\Omega _{\Lambda }}{%
3n^{2}m_{p}^{2}+\gamma \eta ^{-2}\ln (m_{p}^{2}\eta ^{2})+\beta \eta ^{-2}}}%
}{3n^{2}m_{p}^{2}\eta ^{-2}+\gamma \eta ^{-4}\ln (m_{p}^{2}\eta ^{2})+\beta
\eta ^{-4}} +3b^{2}(\frac{1+\Omega _{k}}{\Omega _{\Lambda }})}%
\right) ^{1/2},
\end{equation}%
which takes the form%
\begin{equation}
\phi (a)-\phi (a_{0})=\int_{a_0}^{a}\frac{1}{aH}\left(
\frac{4+\frac{4\chi}{3a}\frac{\sqrt{\frac{3m_{p}^{2}\Omega _{\Lambda
}}{3n^{2}m_{p}^{2}+\gamma \eta ^{-2}\ln (m_{p}^{2}\eta ^{2})+\beta
\eta ^{-2}}}}{3n^{2}m_{p}^{2}\eta ^{-2}+\gamma \eta ^{-4}\ln
(m_{p}^{2}\eta ^{2})+\beta \eta ^{-4}}
+2b^{2}(\frac{1+\Omega _{k}}{\Omega _{\Lambda }})}{4+\frac{2\chi}{a}\frac{\sqrt{%
\frac{3m_{p}^{2}\Omega _{\Lambda }}{3n^{2}m_{p}^{2}+\gamma \eta
^{-2}\ln (m_{p}^{2}\eta ^{2})+\beta \eta
^{-2}}}}{3n^{2}m_{p}^{2}\eta ^{-2}+\gamma
\eta ^{-4}\ln (m_{p}^{2}\eta ^{2})+\beta \eta ^{-4}} +3b^{2}(%
\frac{1+\Omega _{k}}{\Omega _{\Lambda }})}\right) ^{1/2}da.
\end{equation}

\subsection{\textbf{Entropy-corrected new agegraphic tachyon model}}

In recent years, a huge interest has been devoted in studying the
inflationary model with the help of tachyon field. The tachyon field
associated with unstable D-branes might be responsible for
cosmological inflation in the early evolution of the universe, due
to tachyon condensation near the top of the effective scalar
potential \cite{23}, which could suggests some new form of dark
matter at late epoch \cite{24}. A rolling tachyon has an interesting
equation of state whose parameter smoothly interpolates between -1
and 0. This leads us to construct viable cosmological models by
taking the tachyon as an appropriate candidate to explain inflation
at high energy \cite{25}. The effective Lagrangian density
of tachyon matter is given by \cite{6}%
\begin{equation}
L=-V(\phi )\sqrt{1+\partial _{\mu }\phi \partial ^{\mu }\phi },
\end{equation}%
where $V(\phi )$ is the tachyon potential. The energy density and pressure
for the tachyon are written to be%
\begin{eqnarray}
\rho _{T} &=&\frac{V(\phi )}{\sqrt{1-\dot{\phi}^{2}}} \\
p_{T} &=&-V(\phi )\sqrt{1-\dot{\phi}^{2}},
\end{eqnarray}%
where $V(\phi )$ represents the tachyon potential. While the equation of
state of the tachyon is given by%
\begin{equation}
\omega _{T}=\frac{p_{T}}{\rho _{T}}=\dot{\phi}^{2}-1.
\end{equation}%
In order to develope the correspondence between the ECNADE and tachyon dark
energy, we compare Eqs.(25) and (14), and obtain%
\begin{equation}
\dot{\phi}^{2}=-\frac{2\chi}{3a}\frac{\sqrt{\frac{3m_{p}^{2}\Omega
_{\Lambda }}{
3n^{2}m_{p}^{2}+\gamma \eta ^{-2}\ln (m_{p}^{2}\eta ^{2})+\beta \eta ^{-2}}}%
}{(3n^{2}m_{p}^{2}\eta ^{-2}+\gamma \eta ^{-4}\ln (m_{p}^{2}\eta
^{2})+\beta \eta ^{-4})} -b^{2}(\frac{1+\Omega _{k}}{\Omega
_{\Lambda }})
\end{equation}%
Now equating the Eqs. (23) and (9), we get the following expression of
potential energy for the tachyon%
\begin{eqnarray}
V(\phi ) &=&(3n^{2}m_{p}^{2}\eta ^{-2}+\gamma \eta ^{-4}\ln (m_{p}^{2}\eta
^{2})+\beta \eta ^{-4})  \nonumber \\
&&\times \Big(1+\frac{2\chi}{3a}\frac{\sqrt{\frac{3m_{p}^{2}\Omega _{\Lambda }}{%
3n^{2}m_{p}^{2}+\gamma \eta ^{-2}\ln (m_{p}^{2}\eta ^{2})+\beta \eta ^{-2}}}%
}{(3n^{2}m_{p}^{2}\eta ^{-2}+\gamma \eta ^{-4}\ln (m_{p}^{2}\eta
^{2})+\beta
\eta ^{-4})}+b^{2}(\frac{1+\Omega _{k}}{\Omega _{\Lambda }}%
)\Big)^{1/2}.
\end{eqnarray}%
We obtain the evolutionary form of the
tachyon scalar field%
\begin{equation}
\phi (a)-\phi (a_{0})=\int_{a_0}^{a}\frac{1}{aH}\left( -\frac{\frac{2\chi}{3a}\sqrt{%
\frac{3m_{p}^{2}\Omega _{\Lambda }}{3n^{2}m_{p}^{2}+\gamma \eta ^{-2}\ln
(m_{p}^{2}\eta ^{2})+\beta \eta ^{-2}}}}{3n^{2}m_{p}^{2}\eta ^{-2}+\gamma
\eta ^{-4}\ln (m_{p}^{2}\eta ^{2})+\beta \eta ^{-4}} -b^{2}(\frac{%
1+\Omega _{k}}{\Omega _{\Lambda }})\right) ^{1/2}da.
\end{equation}%

\subsection{\textbf{Entropy-corrected new agegraphic dilaton model}}

A dilaton scalar field which exhibits the features of dark energy is
usually originated from the lower-energy limit of string theory.
This model is explained by a general four-dimensional effective
low-energy string action. It has been shown \cite{8} that a scalar
field possessing negative kinetic term (usually known as phantom
type scalar field) does not necessarily lead to inconsistencies
provided that one takes an appropriate structure of higher order
kinetic terms in the effective (underlying) theory. The pressure
density and the energy density of the dilaton dark energy model is
given by \cite{8}%
\begin{eqnarray}
p_{D} &=&-X+ce^{\lambda \phi }X^{2}, \\
\rho _{D} &=&-X+3ce^{\lambda \phi }X^{2},
\end{eqnarray}%
where $c$ and $\lambda $ are positive constants and $\dot{\phi}^{2}=2X.$ The
EoS parameter for the dilaton scalar field is given by%
\begin{equation}
\omega _{D}=\frac{p_{D}}{\rho _{D}}=\frac{-1+ce^{\lambda \phi }X}{%
-1+3ce^{\lambda \phi }X}.
\end{equation}%
Following the same steps as done for the above cases, the comparison of Eq.
(31) with (14), yields%
\begin{equation}
ce^{\lambda \phi
}X=\frac{2+\frac{2\chi}{3a}\frac{\sqrt{\frac{3m_{p}^{2}\Omega
_{\Lambda }}{3n^{2}m_{p}^{2}+\gamma \eta ^{-2}\ln (m_{p}^{2}\eta
^{2})+\beta \eta ^{-2}}}}{3n^{2}m_{p}^{2}\eta ^{-2}+\gamma \eta
^{-4}\ln (m_{p}^{2}\eta ^{2})+\beta \eta ^{-4}}
+b^{2}(\frac{1+\Omega _{k}}{\Omega
_{\Lambda }})}{4+\frac{2\chi}{a}\frac{\sqrt{\frac{3m_{p}^{2}\Omega _{\Lambda }}{%
3n^{2}m_{p}^{2}+\gamma \eta ^{-2}\ln (m_{p}^{2}\eta ^{2})+\beta \eta ^{-2}}}%
}{3n^{2}m_{p}^{2}\eta ^{-2}+\gamma \eta ^{-4}\ln (m_{p}^{2}\eta
^{2})+\beta \eta ^{-4}} +3b^{2}(\frac{1+\Omega _{k}}{\Omega
_{\Lambda }})}.
\end{equation}%
Insert the value of $X$ i.e. $X=\dot{\phi}^{2}/2$ in the above
equation (32),
we get%
\begin{equation}
ce^{\lambda \phi }\dot{\phi}^{2}=\frac{4+\frac{4\chi}{3a}\frac{\sqrt{\frac{%
3m_{p}^{2}\Omega _{\Lambda }}{3n^{2}m_{p}^{2}+\gamma \eta ^{-2}\ln
(m_{p}^{2}\eta ^{2})+\beta \eta ^{-2}}}}{3n^{2}m_{p}^{2}\eta
^{-2}+\gamma
\eta ^{-4}\ln (m_{p}^{2}\eta ^{2})+\beta \eta ^{-4}} +2b^{2}(%
\frac{1+\Omega _{k}}{\Omega _{\Lambda }})}{4+\frac{2\chi}{a}\frac{\sqrt{\frac{%
3m_{p}^{2}\Omega _{\Lambda }}{3n^{2}m_{p}^{2}+\gamma \eta ^{-2}\ln
(m_{p}^{2}\eta ^{2})+\beta \eta ^{-2}}}}{3n^{2}m_{p}^{2}\eta
^{-2}+\gamma
\eta ^{-4}\ln (m_{p}^{2}\eta ^{2})+\beta \eta ^{-4}} +3b^{2}(%
\frac{1+\Omega _{k}}{\Omega _{\Lambda }})}.
\end{equation}%
The above equation can be written as%
\begin{equation}
e^{\frac{\lambda \phi }{2}}\dot{\phi}=\left(
\frac{1}{c}\frac{4+\frac{4\chi}{3a}\frac{\sqrt{\frac{3m_{p}^{2}\Omega
_{\Lambda }}{3n^{2}m_{p}^{2}+\gamma \eta ^{-2}\ln (m_{p}^{2}\eta
^{2})+\beta \eta ^{-2}}}}{3n^{2}m_{p}^{2}\eta ^{-2}+\gamma \eta
^{-4}\ln (m_{p}^{2}\eta ^{2})+\beta \eta ^{-4}}
+2b^{2}(\frac{1+\Omega _{k}}{\Omega _{\Lambda }})}{4+\frac{2\chi}{a}\frac{\sqrt{%
\frac{3m_{p}^{2}\Omega _{\Lambda }}{3n^{2}m_{p}^{2}+\gamma \eta
^{-2}\ln (m_{p}^{2}\eta ^{2})+\beta \eta
^{-2}}}}{3n^{2}m_{p}^{2}\eta ^{-2}+\gamma
\eta ^{-4}\ln (m_{p}^{2}\eta ^{2})+\beta \eta ^{-4}} +3b^{2}(%
\frac{1+\Omega _{k}}{\Omega _{\Lambda }})}\right) ^{1/2},
\end{equation}%
and its integration yields%
\begin{equation}
\phi (a)=\frac{2}{\lambda }\ln \left[ e^{\frac{\lambda \phi (a_{0})}{2}}+%
\frac{\lambda }{2\sqrt{c}}\int_{a_0}^{a}\frac{1}{aH}\left(
\frac{4+\frac{4\chi}{3a}\frac{\sqrt{\frac{3m_{p}^{2}\Omega _{\Lambda
}}{3n^{2}m_{p}^{2}+\gamma \eta ^{-2}\ln (m_{p}^{2}\eta ^{2})+\beta
\eta ^{-2}}}}{3n^{2}m_{p}^{2}\eta ^{-2}+\gamma \eta ^{-4}\ln
(m_{p}^{2}\eta ^{2})+\beta \eta ^{-4}}\ +b^{2}(\frac{1+\Omega
_{k}}{\Omega _{\Lambda }})}{4+\frac{2\chi}{a}\frac{\sqrt{
\frac{3m_{p}^{2}\Omega _{\Lambda }}{3n^{2}m_{p}^{2}+\gamma \eta
^{-2}\ln (m_{p}^{2}\eta ^{2})+\beta \eta
^{-2}}}}{3n^{2}m_{p}^{2}\eta ^{-2}+\gamma
\eta ^{-4}\ln (m_{p}^{2}\eta ^{2})+\beta \eta ^{-4}}+b^{2}(\frac{%
1+\Omega _{k}}{\Omega _{\Lambda }})}\right) ^{1/2}da\right] .
\end{equation}
In summary, in this manuscript we have discussed the new agegraphic
dark energy model from the inclusion of quantum correction in the
entropy-area relation. This ECNADE interacts with cold dark matter
in the FRW spacetime. We have established a correspondence between
ECNADE density with different scalar fields namely, tachyon,
K-essence and dilaton scalar field in the non-flat FRW universe. We
have reconstructed the potentials and the kinetic energies
corresponding to each model which describe tachyon, K-essence and
dilaton cosmology.


\begin{thebibliography}{99}
\bibitem{1} S. perlmutter et al., Astrophys. J. \textbf{517} (1999) 565; A.
G.Riess et al., Astron. J. \textbf{116} (1998)1009;

D. N. Spergel et al., Astrophys. J. \textbf{148} (2003) 175; D. N. Spergel
et al., Astrophys. J. \textbf{170} (2007) 377;

E. J. Copland et al., hep-th/0603057.

\bibitem{2} E. J. Steinhardt et al., Princeton University Prss, Princeton U.
S. A.

\bibitem{3} P. J. E. Peebles and B. Ratra, Astro. J. \textbf{325} (1988) $L$%
17; C. Wetterich, Nucl. Phys. \textbf{B 302} (1988) 668;

P.J.Steinhardt, Phys. Rev. Lett. \textbf{82 }(1999) 896; I. Zlativ et al.,
Phys. Rev. Lett. \textbf{82} (1999) 82.

\bibitem{4} E. Elizalde et al., Phys. Rev. \textbf{D} \textbf{70} (2004)
043539; S. Nojiri et al., Phys. Rev. \textbf{D 71 }(2005) 063004; A.
Anisimov et al., J. Cosmol. Astropart. Phys. \textbf{06} (2005) 006.

\bibitem{5} T. Chiba et al., Phys. Rev. \textbf{D 62} (2000) 023511; C.
Armendariz-Picon et al., Phys. Rev. Lett \textbf{85} (2000) 4438;

C. Armendariz-Picon et al., Phys. Rev. Lett \textbf{63} (2001) 103510.

\bibitem{6} A. Sen, JHEP \textbf{10} (1999) 008; T. R. Choudhary, Phys. Rev.
\textbf{D 66} (2002) 081301; E. A. Bergshoeff et al., JHEP \textbf{05}
(2000) 009; L. R. W. Abramo, F. Finelli, Phys. Lett. \textbf{B 575} (2003)
165.

\bibitem{7} R. R. Caldwell, Phys. Lett. \textbf{B 545} (2002) 23; S. Nojiri,
S. D. Odinstov, Phys. Lett. \textbf{B 562 }(2003) 147; S. Nojiri, S. D.
Odinstov, Phys. Lett. \textbf{B 565 }(2003) 1.

\bibitem{8} M. Gasperini et al., Phys. Rev. \textbf{D 65} (2002) 023508; N.
Arkani-Hamed et al., J. Cosmol. astropart. Phys. \textbf{04} (2004) 001; F.
Piazza, S. Tsujikawa, J. Cosmol. Astropart. Phys. \textbf{07} (2004) 004.

\bibitem{9} M. Jmail, M. A. Rashid, Eur. Phys. J. \textbf{C 60} (2009) 141;
M Jamil, F. Rahman, Eur. Phys. J. \textbf{C 64} (2009) 97; W. Zimdahl, Int.
J. Mod. Phys. \textbf{D 14} (2005) 2319; H. Zhang et al., Phys. Lett.
\textbf{B 678} (2009) 331.

\bibitem{10} C. Daffayet et al., Phys Rev. \textbf{D 65} (2002) 044023; Y.
Shtanov, j. Cosmol. Astropart. Phys. \textbf{11} (2003) 014.

\bibitem{11} A. Kamenshchik et al., Phys. Lett. \textbf{B 511} (2001) 511;
M. C. Bento et al., Phys. Rev. \textbf{D 66} (2002) 043507.

\bibitem{12} S. D. Hsu, Phys. Lett. \textbf{B 594} (2004) 594; M. Li, Phys.
Lett. \textbf{B 603} (2004) 1.

\bibitem{13} R. G. Cai, Phys. Lett. \textbf{B 657} (2007) 228.

\bibitem{14} G't Hooft, preprint gr-qc/ 9310026 (1993).

\bibitem{15} A. G. Cohen et al., Phys. Rev. Lett. \textbf{82} (1999) 4971.

\bibitem{16} T. Zhu, J. R. Ren Eur. Phys. J. \textbf{C 62} (2009) 413; R. G.
Cai et al., Class. Quant. Grav. \textbf{26} ( 2009) 155018.

\bibitem{17} C. Rovelli, Phys. Rev. Lett. \textbf{77} (1996) 3288; A
Ashtekar, et al., Phys. Rev. Lett. \textbf{80} (1998) 904; A. Ghosh, P.
Mitra, Phys. Rev. \textbf{D 71} (2005) 027502; K. A. Meissner, Class. Quant.
Grav.\textbf{\ 21} (2004) 5245.

\bibitem{18} H. Wei, Commun. Theor. Phys. \textbf{52} (2009) 743; M. Jamil,
M. U. Farooq JCAP \textbf{03} (2010) 001.

\bibitem{19} H. Wei, S. N. Zhang, arXiv: 0707.2129 [astro-ph].

\bibitem{20} F. Karolyhazy, Nuov Cim. \textbf{A 42} (1966) 390; F.
Karolyhazy et al., \textit{in Physics as Natural Philosophy, }edited by A.
Simony and H. feschbach, MIT Press, Cambridge, MA (1982); F. Karolyhazy et
al., \textit{in Quantum concepts in Space and Time}, edited by R. Penrose
and C. J. Isham, Clarendon Press, Oxford (1986).

\bibitem{21} H. Wei, R. G. Cai, Phys. Lett. \textbf{B} \textbf{660} (2008)
1; arXiv: 0708.0884.

\bibitem{22} H. Wei, R. G. Cai, Phys. Lett. \textbf{B 663} (2008) 1; arXiv:
0708.1894.

\bibitem{23} A. Sen, Mod. Phys. Lett. \textbf{A 17} (2002) 1797; N. D.
Lambert, I. Sachs, Phys. Rev. \textbf{D 67} (2003) 026005.

\bibitem{24} M. Sami et al., Phys. Rev. \textbf{D 66} (2002) 043530.

\bibitem{25} A. Mazumdar et al., Nucl. Phys. \textbf{B 614} (2001) 101; M.
Fairbairn, M. H. G. Tytgat, Phys. Lett. \textbf{B 546} (2002) 1; M. Sami,
Mod. Phys. Lett. \textbf{A 18} (2003) 691; Y. S. Piao et al., Phys. Rev.
\textbf{D 66} (2002) 121301.
\end{thebibliography}
\end{document}